\documentclass[runningheads]{llncs}

\usepackage[utf8]{inputenc} 

\usepackage{graphicx}
\usepackage{amsmath,amssymb}

\newcommand{\FL}{\mathbf{L}}
\newcommand{\FM}{\mathbf{M}}
\newcommand{\FZ}{\mathbf{0}}
\newcommand{\FOne}{\mathbf{1}}
\newcommand{\Fx}{\mathbf{x}}
\newcommand{\Fm}{\mathbf{m}}

\begin{document}
\title{Learning the GENERIC evolution\thanks{Supported by the Czech Science Foundation, Project No. 20-22092S.}}
%
%
\author{Martin Šípka\inst{1}\orcidID{0000-0002-7176-7538} \and
Michal Pavelka\inst{1}\orcidID{0000-0003-0605-6737}}
\authorrunning{M. Šípka et al.}
%
\institute{Mathematical Institute, Faculty of Mathematics and Physics, Charles University, \\
Sokolovská 83, 186 75 Prague, Czech Republic.
\email{martinsipka@gmail.com}}
\maketitle              
\begin{abstract}
We propose a novel approach for learning the evolution that employs differentiable neural networks to approximate the full GENERIC structure. Instead of manually choosing the fitted parameters, we learn the whole model together with the evolution equations. We can reconstruct the energy and entropy functions for the system under various assumptions and accurately capture systems behaviour for a double thermoelastic pendulum and a rigid body.

\end{abstract}
\section{Introduction}

When analyzing the behaviour of observed systems, it is natural to try and fit a model to describe them. Depending on the complexity of the phenomena present, we may need to consider different assumptions to adequately capture the nature and character of our problems. Probably the most common approach is to consider evolutionary equations known, yet identify model parameters, which can be for example viscosity in the case of fluids or moments of inertia for rigid bodies. Next, we use the assumed equations of the system and available experimental data to fit these parameters. After such a process, we consider the system described, and we may proceed with the simulations and predict the outcome of unseen events. In this paper, we consider a different approach. We try to solve the problem of what to do when the equations are only partially available or even completely unknown. We will employ the framework named General Equation for Non-Equilibrium Reversible-Irreversible Coupling (GENERIC) \cite{generic} and use it to fit also evolutionary equations for the observed data. The minimalistic assumptions of the framework will serve as a guide for the training process and will result in thermodynamically consistent simulation. To approximate functions and their derivatives, we choose to work with deep neural networks, as by employing them, we can, in theory, train arbitrarily complex evolution.

The novelty of the paper lies in the following. To our best knowledge, full reconstruction of the four main components of the GENERIC evolution was never attempted before. Obtaining full energy and entropy functions together with a Poisson bivector $\FL$ and a friction matrix $\FM$, compliant with basic thermodynamics principles is a new approach to dynamic system fitting.
\section{Model}
We prescribe a generic structure in the form
\begin{equation}
\dot{\mathbf{x}} = \FL(\mathbf{x}) \cdot \nabla E(\Fx) + \FM(\Fx) \cdot \nabla S(\Fx).
\end{equation}
We will be dealing only with discrete systems, therefore the derivatives will always be simple partial derivatives. To ensure the basic principles of energy conservation and entropy production we construct our Poisson bivector $\FL$ to be antisymmetric and the friction matrix $\FM$ to be symmetric. Moreover we need to fulfill the so-called degeneracy conditions
\begin{subequations}  \label{reg} 
\begin{align} 
\FL(\Fx) \cdot \nabla S(\Fx) &= \FZ \label{reg:Entropy}\\
\FM(\Fx) \cdot \nabla E(\Fx) &= \FZ \label{reg:Energy}.
\end{align}
\end{subequations}
There is one more condition for $\FL$ that needs to hold. As $\FL$ represents a Poisson bracket, the Jacobi identity
\begin{equation} \label{jacobi}
L^{kl}\partial_k L^{ij} + L^{kj} \partial_k L^{li} + L^{ki} \partial_k L^{jl} = 0,
\end{equation}
is a requirement. To create the GENERIC structure in our code, we construct the neural networks in the following way.
\subsubsection{Energy} is a scalar output, fully connected neural network with all the state variables as inputs. We construct the network such that $E \in C^\infty$. This can be achieved by the proper activation function.
\subsubsection{Entropy} has the same structure as energy. 
\subsubsection{L,} the Poisson bivector, is constructed as a fully connected neural network with state variables as inputs and $n(n+1)/2$ outputs. These outputs are then rearranged so they create an antisymmetric matrix as prescribed by generic evolution. Note that $n$ of the elements are not needed and we would save some computational resources by outputting only $n(n-1)/2$ elements but we have not yet implemented such a mechanism in Tensorflow.
\subsubsection{M,} the friction matrix, has the same structure as $\FL$, except it is symmetric and uses all $n(n+1)/2$ elements. $\FM$ is required to be positive semi-definite, therefore $\sqrt{\FM}$ is predicted by the network and the output is then multiplied with itself. 
\newline \newline
We simulated the described system and collected the training data. We worked with the Crank-Nicholson scheme as it is the suitable integrator for double pendulum and rigid body problems. 
\section{Training}
Training is done using Tensorflow framework. We first simulate a set of trajectories and save the simulated points in pairs $(x_n, x_{n+1})$. Next we define a loss to minimize. The total loss of one step is composed of three parts. The first component is the trajectory prediction
\begin{align}
\mathcal{L}_{traj} = \frac{2(\Fx_n - \Fx_{n+1})}{dt} + \FL(\Fx_n) \cdot \nabla E(\Fx_n) + \FM(\Fx_n) \cdot \nabla S(\Fx_n) + \\ \FL(\Fx_{n+1}) \cdot \nabla E(\Fx_{n+1}) + \FM(\Fx_{n+1}) \cdot \nabla S(\Fx_{n+1}),
\end{align}
which is just a Crank-Nicholson scheme rewritten, so it equals zero for the exact trajectory. Note that we assume the timestep $dt$ to be known, yet not necessarily constant, as we usually have acces to a time measuring mechanism while observing a physical system. The second part of the loss follows directly from~\eqref{reg}
\begin{subequations}
\begin{align}
\mathcal{L}_{regS} &= \FL(\Fx_n) \cdot \nabla S(\Fx_n) \\
\mathcal{L}_{regE} &= \FM(\Fx_n) \cdot \nabla E(\Fx_n).
\end{align}
\end{subequations}
The last part of the loss represents the Jacobi identity, and follows directly from~\eqref{jacobi}.
\begin{equation}
\mathcal{L}_{Jac} = L^{kl}\partial_k L^{ij} + L^{kj} \partial_k L^{li} + L^{ki} \partial_k L^{jl}.
\end{equation}
Together the loss is constructed as
\begin{equation}
\mathcal{L} = w_t\mathcal{L}_{traj}^2 + w_S\mathcal{L}_{regE}^2 +w_L\mathcal{L}_{regS}^2 + w_J\mathcal{L}_{Jac}^2.
\end{equation}
The last three terms in the loss function are vectors or tensors. In order to make the loss scalar, one simply squares the components and then sums them. $w$ values are weights to allow greater flexibility and are considered hyperparameters of the method. Since all the operations are differentiable, we can backpropagate the loss and use standard gradient descend based approaches. Gradients of the energy and the entropy and the gradients in Jacobi identity are constructed from the networks using a standard automatic differentiation package in Tensorflow. Note that the Jacobi identity-based loss is not yet used in the first version of the paper.

\section{Results}
We illustrate the approach on two systems. First, we consider a double elastic pendulum taking inspiration from \cite{romero} and we work in energetic representation. This system is also learned in \cite{quercus} in a similar way. We, however, use a different set of assumptions. We assume to know the physical meaning of the observed variables, therefore we know that entropy is a trivial sum of particle entropies and that $\FL$ is canonical. The $\FM$ and $E(\Fx)$ is learned. 

Next, we proceed to learn the non-dissipative rigid body dynamics and its energy. This is to illustrate the learning of systems with non-canonical Poisson brackets without any apriori assumptions outside the standard GENERIC framework. 

Lastly, we study the double pendulum once again, but we observe temperature instead of entropy. We, therefore, do not know any of the $\FL, \FM, E, S$ and we need to learn them all from data. This situation can happen also in practice, either when we know the variables we are observing but we have no information about the evolution, or when we are measuring a set of data of unknown physical meaning. In the latter case, by training the model we might even verify that a GENERIC structure exists for a given set of data by studying the training error.

\subsection{Double pendulum in energetic representation}
Let us first consider a problem from the paper \cite{romero}. We work with a simple system. Two masses $m_i$ are connected by springs parametrized by spring constants $C_i$. The springs have its current length $\lambda_i$ and its natural length $\lambda_i^0$ Each spring has also its entropy $s_i$ and therefore a temperature $\theta_i$. The reference temperature is denoted as $\theta_{ref}$. The energy of the system is chosen as:
\begin{equation}
\begin{split}
E(\mathbf{q}, \mathbf{p}, s)  = \sum_i \frac{\mathbf{p}_i^2}{2m_1} + \frac{C_i}{2}\log^2 \frac{\lambda_i}{\lambda_i^0} + \beta \theta_{ref} \log \frac{\lambda_i}{\lambda_i^0} + \\ c_0 \theta_{ref}\Big(\exp \Big(\frac{s_i-\beta \log (\lambda_i/\lambda_i^0)}{c_0}\Big) - 1\Big).
\end{split}
\end{equation}
We can notice that the mechanical and thermal evolution is coupled. This coupling is expressed by a parameter $\beta$. Entropy is in this representation trivial
\begin{equation}
S(\mathbf{q}, \mathbf{p}, s) =  \sum_i s_i
\end{equation}
$\FL$ in this representation reads
\begin{equation}
\FL = \begin{pmatrix}
\FZ & \FZ & \FOne \ & \FZ \ & \FZ \ & \FZ\\
\FZ & \FZ & \FZ & \FOne & \FZ & \FZ\\
-\FOne & \FZ & \FZ & \FZ & \FZ & \FZ\\
\FZ & -\FOne & \FZ & \FZ & \FZ & \FZ\\
\FZ & \FZ & \FZ & \FZ & 0 & 0\\
\FZ & \FZ & \FZ & \FZ & 0 & 0\\
\end{pmatrix}.
\end{equation} 
\subsubsection{Training of the pendulum}
Since $\FL(\Fx)$ and the entropy is trivial, we restrict ourselves to train only functions $\FM(\Fx)$ and $E(\Fx)$. The Jacobi identity and \eqref{reg:Entropy} is fulfilled automatically for our case. The loss function, therefore, simplifies to  
\begin{equation}
\mathcal{L} = w_t\mathcal{L}_{traj}^2 +w_L\mathcal{L}_{regE}^2
\end{equation}
For this problem we modified a neural network design by featurising input to $(\mathbf{q}_1^2, (\mathbf{q}_2 - \mathbf{q}_1)^2, \mathbf{p}_1^2, \mathbf{p}_2^2)$ in order to speed up the training. We could let the network learn this representation but we would generally need more data and computational resources. After 4 iterations the training converged and we could run the simulations. The system clearly behaved as a double pendulum. Error in the simulations could be decreased by using more data and more sophisticated neural networks. It is interesting to note that the learned energy was conserved up to an error of 1\%. The value of the learned energy was not close to the original value but this is caused by only using its gradient during training and simulations.
\begin{figure}
\includegraphics[width=\textwidth]{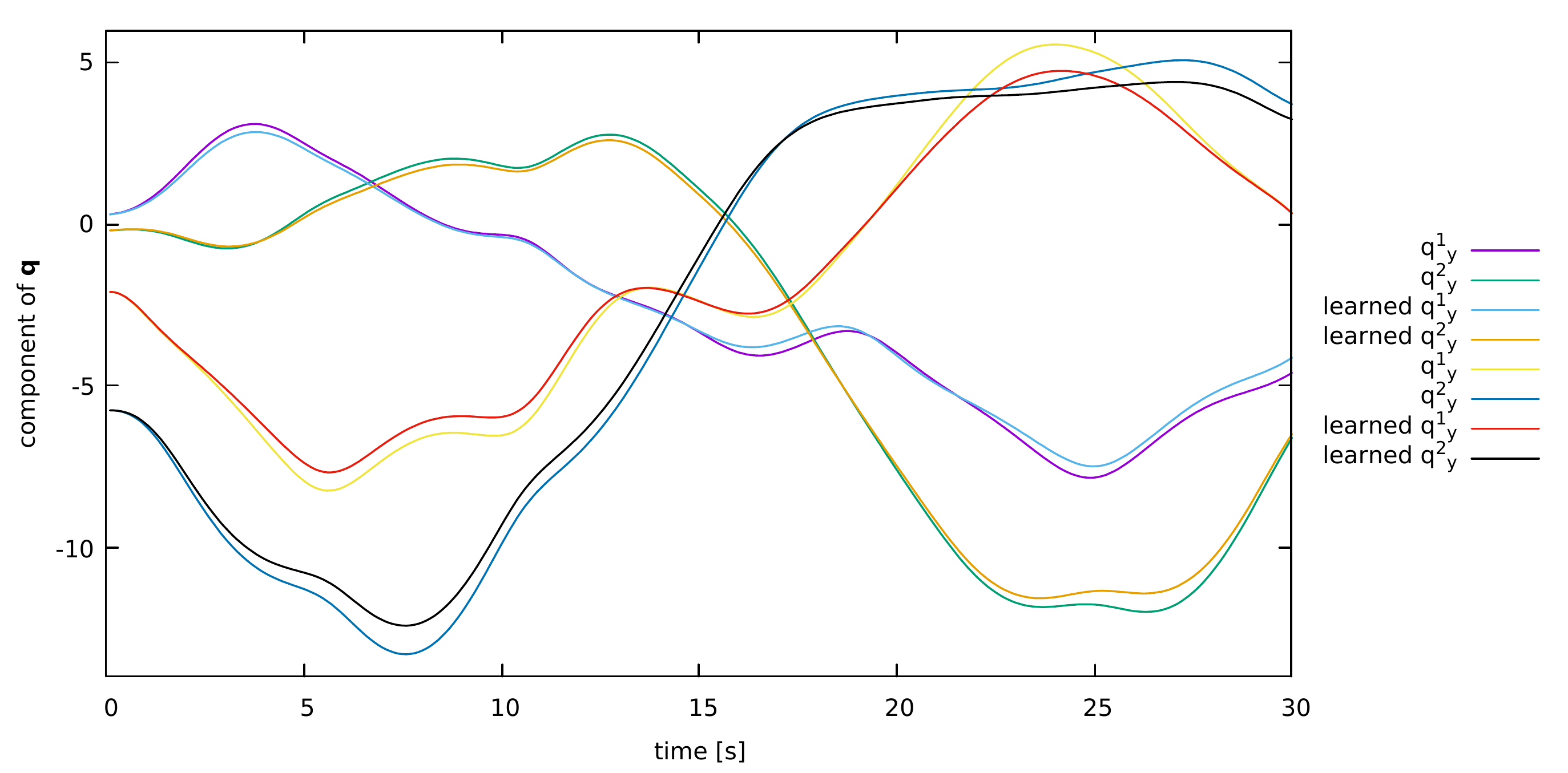}
\caption{Possitions of the two masses during the evolution of double thermoelastic pendulum. The learned trajectory matches the original one quite well. We simulated from the initial conditions not seen in the training data.} \label{fig1}
\end{figure}
\subsection{Rigid body dynamics}
\begin{figure}
\includegraphics[width=\textwidth]{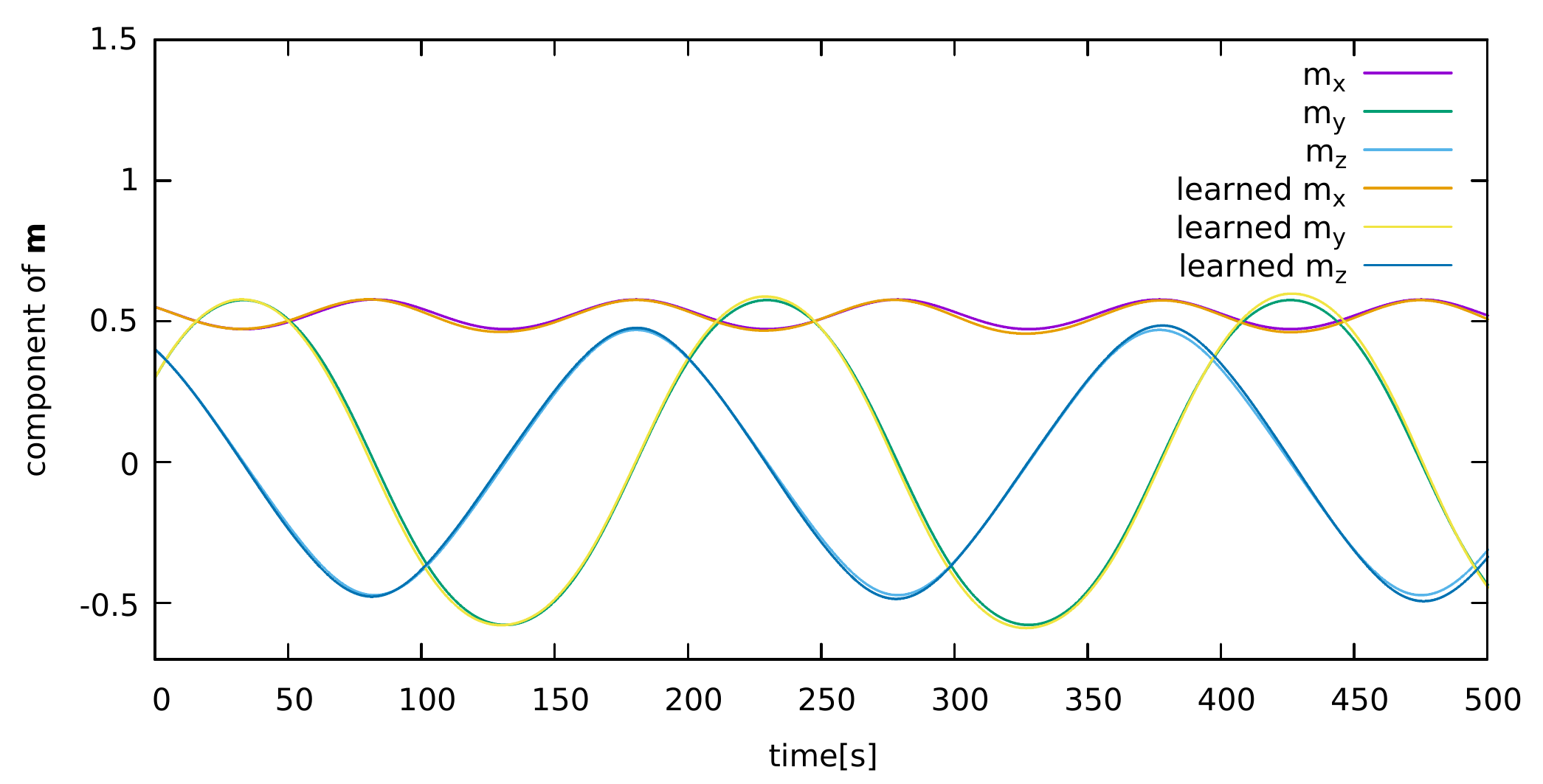}
\caption{Rigid body trajectory simulated from exact dynamics compared to the learned evolution. Initial conditions were chosen to be different than in the training trajectories.} \label{rigid_traj}
\end{figure}

To better understand a training of a simple but noncanonical Poisson bracket we simulate the dynamics of a rigid body. For a better insight on how such a system is described in the GENERIC setting, we refer a reader to \cite{gen_book}. The resulting Poisson bracket is 
\begin{equation}
\FL_{ij} = -\epsilon_{ijk} \Fm_k
\end{equation}
And the energy is quadratic in $\Fm$
\begin{equation}
E(\Fm) = \frac{1}{2} \Fm \cdot \mathbf{I} \ \Fm,
\end{equation}
where $\mathbf{I}$ is the inertia tensor. As in the previous section we simulated 5 trajectories, each 10 000 steps long and then trained the networks. This time, however, we only trained $\FL$ and $E(\Fm)$. The loss function was composed of only the trajectory loss. The Jacobi identity loss should and will be included in the future versions of the paper.
\begin{equation}
\mathcal{L} = w_t\mathcal{L}_{traj}^2 + w_J\mathcal{L}_{Jac}^2
\end{equation}
We split the data so that 80\% was used for training and 20\% for validation. Both training and validation loss quickly vanished.  The training converged and the trajectories well agree with the ones from the exact simulations (see~\ref{rigid_traj}). 
\subsection{Double pendulum in temperature variables}
\begin{figure}
\includegraphics[width=\textwidth]{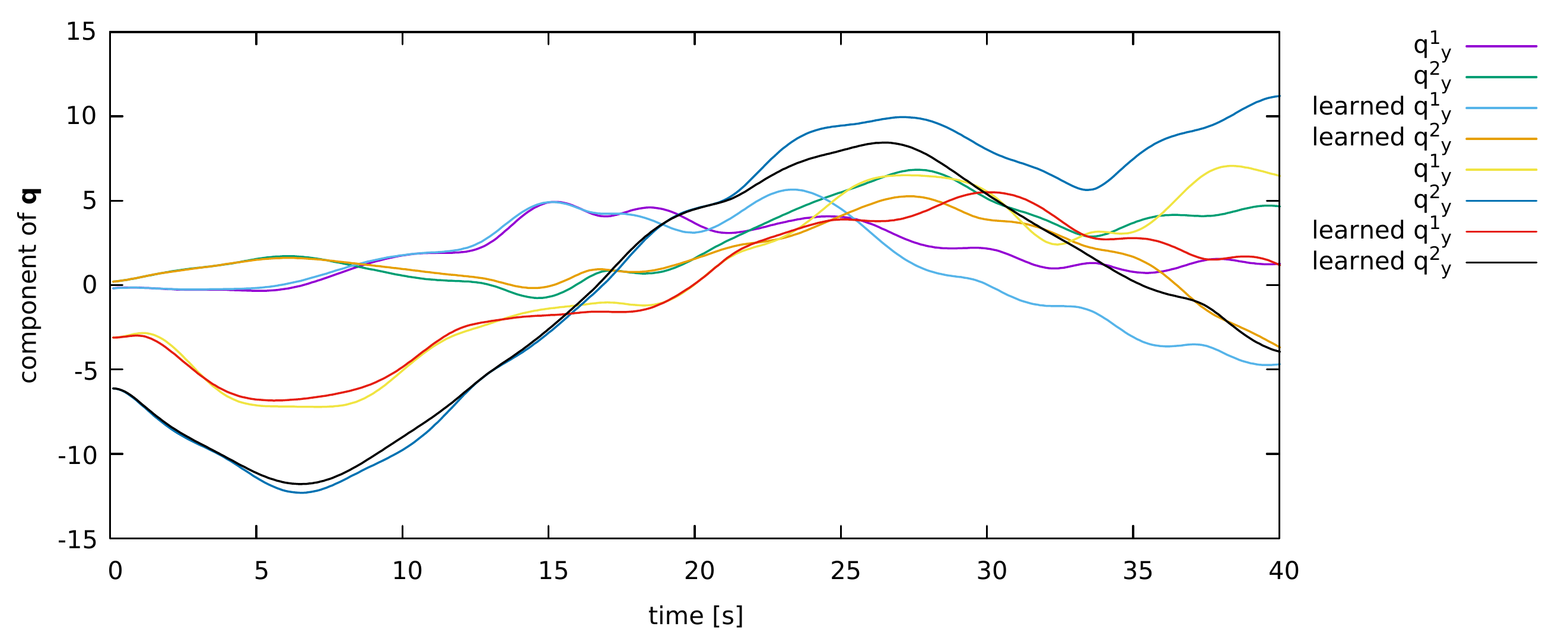}
\caption{Possitions of the two masses during the evolution of a thermoelastic pendulum in $(\mathbf{q}, \mathbf{p}, \theta)$. Initially, the trajectories are similar, however, they diverge as errors accumulate. We simulated from the initial conditions not seen in the training data.} \label{temp}
\end{figure}

From the energetic representation, we now switch to a new set of variables $\Fx = (\mathbf{q}, \mathbf{p}, \theta)$. In this case, $\FL$ and $\FM$ are assumed not known. We should be able to derive them but in the original paper~\cite{romero}, they are described as cumbersome so we will not attempt to. Instead, we consider the most general approach and learn $\FL, \FM, E(\Fx)$ and $S(\Fx)$ from data directly. We again use the featurisation of the input $(\mathbf{q}_1^2, (\mathbf{q}_2 - \mathbf{q}_1)^2, \mathbf{p}_1^2, \mathbf{p}_2^2)$ to speed up the training. After simulating 40 trajectories, each 1000 steps long, the training itself is performed. The system is more difficult to train than for previous problems. However, we are able to achieve convergence and compare the learned trajectory with the simulated one for unknown initial conditions (see~Figure~\ref{temp}). Notice that the evolution follows the exact one at the beginning but as the error accumulates in this chaotical system, we diverge from the solution eventually. The energy value predicted by the neural network was conserved during the evolution almost exactly. 
\section{Related work}
Using machine learning and neural networks in particular to either speed up the simulations or recognise the unknown dynamics has gained popularity in the past years. With the rapid improvement of deep learning, more general and scalable learning techniques emerged also in physics. 
\subsubsection{Learning the Hamiltonian}
In the case of a canonical Poisson bracket and a nondissipative system, it is sufficient to learn only the Hamiltonian function. This was demonstrated by \cite{ham_proc}. Our approach is, however not limited to canonical evolution and can be used also in case there is dissipation in the system. 
\subsubsection{Learning the forcefields}
In Ab-Initio molecular dynamics, neural networks are used to approximate force fields arising from accurate, yet costly Density Functional Theory calculations. This area is actively researched and neural network design and efficiency are improved continuously. See review \cite{reviewMLP} for comprehensive summary. In this case, the equations of motion are known, as we employ the Born-Oppenheimer approximation. Our method offers a solution for when such simplification would not hold and we need to learn more complex dynamics. 
\subsubsection{Neural PDE solver}
One may look at the part of the learning as a partial differential equation (PDE) solver. As we employ the gradient of the neural network representing energy and reconstruct it using dynamics data, we solve the equation
\begin{equation}
\nabla E = \mathbf{f},
\end{equation}
where $\mathbf{f}$ is obtained from data. Using a neural network approximator to solve PDE became an interesting field of research recently. See for example \cite{PDESolver}.
\section{Conclusion}
We demonstrated the learning of evolutionary equations for two different systems with varying assumptions on the character of the GENERIC evolution. The approach is general enough that event without any apriori knowledge of the system, we are able to reconstruct its equation of motion, together with its energy and entropy. All of the quantities are functions of the observed variables modelled by the neural networks. 
The method can be applied in case we are dealing with an unknown system to find the GENERIC evolution and confirm the system is GENERIC compatible. 
\section{Acknowledgments}
We are grateful to  Elías Cueto,  Miroslav Grmela, Francisco Chinesta and Beatriz Moya for interesting discussions about the subject. This work was supported by the Czech Science Foundation, Project No. 20-22092S.
%
%
%
\bibliographystyle{splncs04}
%

\end{document}